\documentclass[11pt,twoside]{article}
\usepackage{asp2010}

\resetcounters

\begin{document}

\title{Evolution of Giant Molecular Clouds in Nearby Galaxies}
\author{Jin Koda$^1$
\affil{$^1$Stony Brook University}}

\begin{abstract}
Our knowledge of GMC evolution in galactic disks has advanced tremendously in past several years.
Studies were limited to local, predominantly atom-rich small galaxies, but have now been expanded
to typical spiral galaxies with a rich molecular content.
The evolution appears quite different between the two environments.
GMCs exist almost exclusively along HI spiral arms and filaments in the disks of
local small galaxies (LMC, M33),
suggesting that GMCs form and end their short lives there.
However, in a more molecular-rich environment
(MW, M51), GMCs are present everywhere independent of HI structures.
Indeed, the molecular gas fraction remains high and almost constant during arm passage
into the next inter-arm region.
The gas remains molecular, presumably in GMCs, for a long time.
A transitional case has been found recently in the central region of the atom-rich galaxy M33 
- GMCs do not coincide with HI there.
Evolution of the physical conditions of molecular gas from spiral arms to inter-arm regions
is also being revealed in molecule-rich galaxies.
An increase of the CO $J=$2-1 and 1-0 line ratio in spiral arms in M51
suggests density and/or temperature increases by a factor of 2-3 in GMCs in the arms,
compared to their counterparts in the inter-arm regions.
An analysis of high-resolution Milky Way survey data revealed
that the fraction of dense (or warm) clumps increases dramatically in the spiral arms.
\end{abstract}

The formation, evolution, and lifetime of GMCs in galaxies are of critical importance to
our understanding of interstellar matter (ISM) and star formation.
The ISM evolves during galactic rotation. Therefore knowing
the distribution of GMCs in galactic disks and their relations to galactic structures
is essential.
This presentation was given at a celebration of the 30th anniversary of
the Nobeyama Radio Observatory (NRO).
I summarize recent progress in the field of GMC evolution with a particular emphasis
on recent results from NRO.

\section{Brief History}

\articlefigure[width=1.0\textwidth]{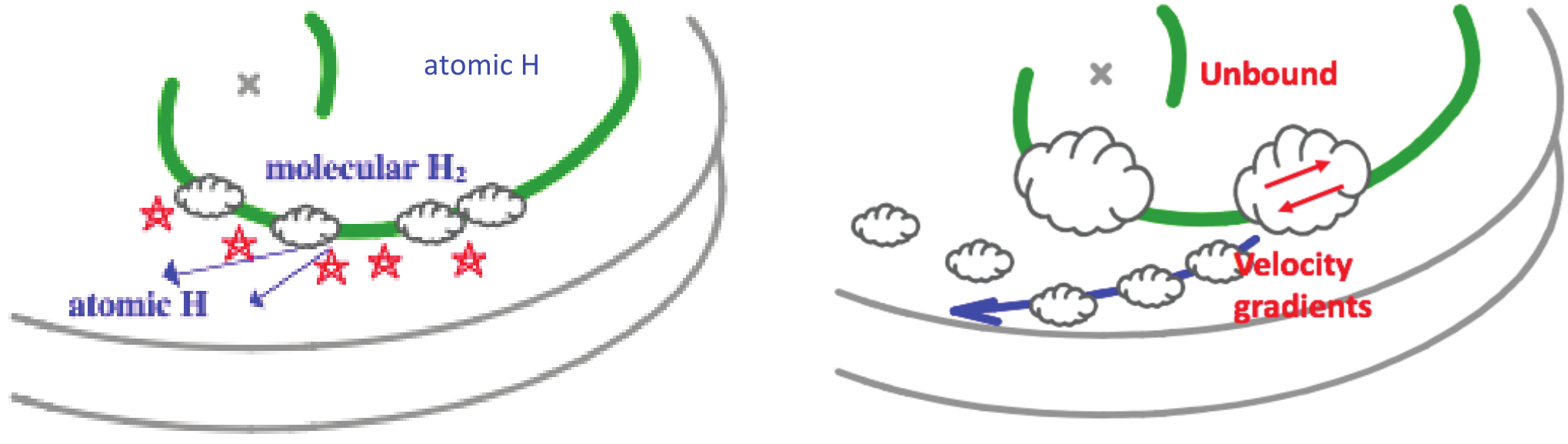}{fig:scheme}{Schematic illustrations of GMC evolution. {\it Left:} Previous picture. {\it Right:} Emerging picture.}

The textbook picture of ISM phases posits that
GMCs are assembled in spiral arm shocks from
diffuse interarm HI gas and then photo-dissociated back into the atomic phase
by OB star formation within the spiral arms (Figure \ref{fig:scheme} {\it left}).
This picture predicts a rapid gas-phase change across spiral arms -- from atomic
to molecular and back into atomic after spiral arm passage.

In retrospect, observational support for this standard picture left some unanswered questions.
There were intense debates on the distribution and lifetime of GMCs in the MW
in 1980's, which were never fully resolved.
Indeed, two major Galactic plane CO ($J$=1-0) surveys arrived at contradictory results.
The Columbia survey (and the Harvard survey later) found very little CO emission in the inter-arm regions
in the longitude-velocity ($l$-$v$) diagram, suggesting that GMCs survive only for
the duration of spiral arm passage \citep[an order of $\sim 20$-$30$ Myr; ][]{Cohen:1980ve, Dame:2001gs}.
On the other hand, the Massachusett-Stony Brook survey found
an abundant population of GMCs even in the inter-arm regions, concluding that GMCs live for the order of
the Galactic rotation timescale \citep[$\gtrsim 100$ Myr; ][]{Sanders:1985ud, Scoville:2004lr}.

Early interferometric observations found only few GMCs in the inter-arm regions
in the grand-design spiral galaxy M51 \citep{Rand:1990fb},
apparently supporting the textbook picture.
However, interferometric observations have a well-known disadvantage in studies of extended nearby galaxies.
Arrays with $\lesssim10$ antennas (e.g., Nobeyama, OVRO, BIMA, and PdBI arrays)
miss most of spatial information, especially in extended inter-arm regions,
being biased toward confined emission along the spiral arms.
Mapping GMCs in inter-arm regions, especially in presence of bright spiral arms,
was difficult even when sensitivity was not an issue.

\section{GMC Distribution and Lifetime}

\subsection{Molecule-rich Galaxies}

A combination of modern arrays with $\geq 15$ antennas and single-dish
telescopes permits high-quality molecular gas imaging.
An abundant GMC population in the inter-arm regions was found in the molecule-rich
galaxy M51, using the Combined Array for Research in Millimeter Astronomy (CARMA)
and the Nobeyama Radio Observatory 45m (NRO45) telescope \citep{Koda:2009wd}.
The combination of CARMA and NRO45 provided an unprecedented high-image quality,
reconstructing the full emission including the most extended component \citep{Koda:2011nx}.

\articlefigure[width=1.0\textwidth]{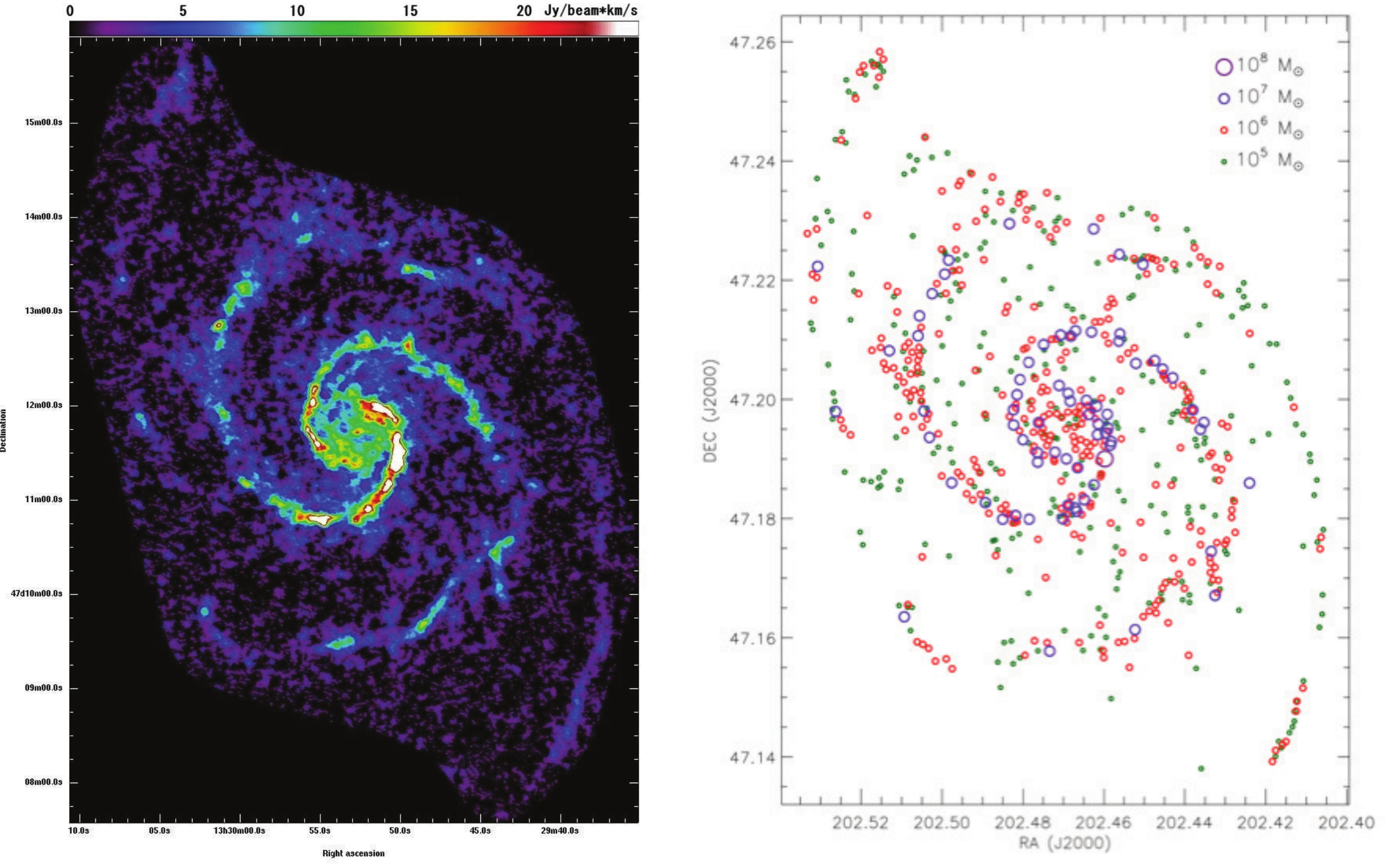}{fig:m51gmc}{CO(1-0) map ({\it left}) and distribution of GMCs ({\it right}) in M51,
based on CARMA and the Nobeyama 45m telescope observations \citep{Koda:2009wd}.}

In M51, the majority of the gas remains molecular from arm entry through the interarm region and
into the next spiral arm passage.
The molecular gas fraction varies only little azimuthally.
The most massive GMCs (giant molecular associations -- GMAs) appear only along spiral arms (Figure \ref{fig:m51gmc}),
suggesting that they are first assembled from pre-existing smaller GMCs entering the spiral arms,
and then broken up as the gas flows through the arms.
The GMAs and their H$_2$ molecules are not fully dissociated into atomic gas as predicted
in stellar feedback scenarios, but are fragmented into smaller GMCs upon leaving the spiral arms.
The remnants of GMAs are detected as the chains of GMCs that emerge from the spiral arms
into interarm regions.
The observed kinematic shear within the spiral arms is sufficient to unbind the GMAs against self-gravity.
\citet{Koda:2009wd} discussed that the evolution of GMCs is driven by large-scale galactic dynamics
-- their coagulation into GMAs is due to spiral arm streaming motions, and
their fragmentation as they leave the arms is due to shear. Therefore, the lifetime of molecular gas
appears to be an order of the Galactic rotation timescale ($\gtrsim 100$ Myr),
while GMCs can be coagulated and/or fragmented to the next-generation GMCs without
passing through the atomic phase.

A similar GMC mass segregation between spiral arms and inter-arm regions is found
in the Milky Way, another molecule-rich galaxy.
GMCs exist also in the inter-arm regions, but their masses are about an order of magnitude smaller
than their counterparts in the spiral arms  \citep{Koda:2006fk}.
The same tend is being found in other molecule-rich spiral galaxies
in the CARMA-Nobeyama Nearby-galaxies (CANON)
CO($J$=1-0) survey, in which 29 local spiral galaxies are observed with both CARMA and NRO45.

\subsection{Atom-rich Galaxies}

Comparisons of molecule-rich and atom-rich galaxies provide another clue
in understanding GMC evolution.
Smaller local (dwarf) galaxies rich in atomic gas (e.g., LMC and M33) show
much fewer GMCs in their disks \citep{Fukui:2009lr, Engargiola:2003jo}.
The GMCs are almost exclusively associated with HI spiral arms and filaments and
are mostly absent in the inter-arm regions.
These observations immediately indicate their short lifetimes
(an arm-crossing timescale of $\sim 30$ Myr).

\citet{Kawamura:2009lr} found a similar lifetime of 20-30 Myr in LMC,
by translating the fractions of GMCs with and without young star clusters
to their lifetimes using cluster ages as a gauge.
\citet{Miura:2012yq} also found a similar lifetime of 20-40 Myr in M33
by comparing CO($J$=3-2) data and star clusters.
The short lifetime of GMCs appears to be common in these atom-rich galaxies.
In retrospect, the observations of extragalactic GMCs beyond the MW were limited
to these closest, predominantly atom-rich galaxies; this bias might have
contributed to the textbook picture of GMC evolution.

As a side note, these relatively-short lifetimes may still be at odds with the even shorter
dissipation timescale of internal velocity dispersions \cite[a few Myr; ][]{Mac-Low:1999rw}.
It is difficult to maintain the velocity dispersions even for the relatively short lifetime
of 20-30 Myr .

\subsection{A Transitional Case}
An interesting transitional case has been found recently in the central 4 kpc region of
M33 \citep[][ see Figure \ref{fig:m33_gmc}]{Tosaki:2011fk}.
In this atom-rich galaxy, GMCs are located predominantly in HI spiral structures
in most of the disk \citep{Engargiola:2003jo}, indicating that GMCs form and die there.
However, they do not coincide with HI structures in the central $r<2$ kpc,
just as in the disks of molecule-rich galaxies.
GMCs are decoupled from the HI distribution as if they are entities that survive
through almost a galactic rotation period.

The molecular gas fraction increases in this central region, although the dominant
phase of gas there is still atomic \citep{Tosaki:2011fk}. There are many parameters
potentially responsible for this transition, including the amount of gas, stellar gravitational potential,
metallicity, radiation field, dynamical environment (such as shear), etc.
Obviously, a much larger sample is necessary to isolate the cause.
 
\articlefigure[width=0.5\textwidth]{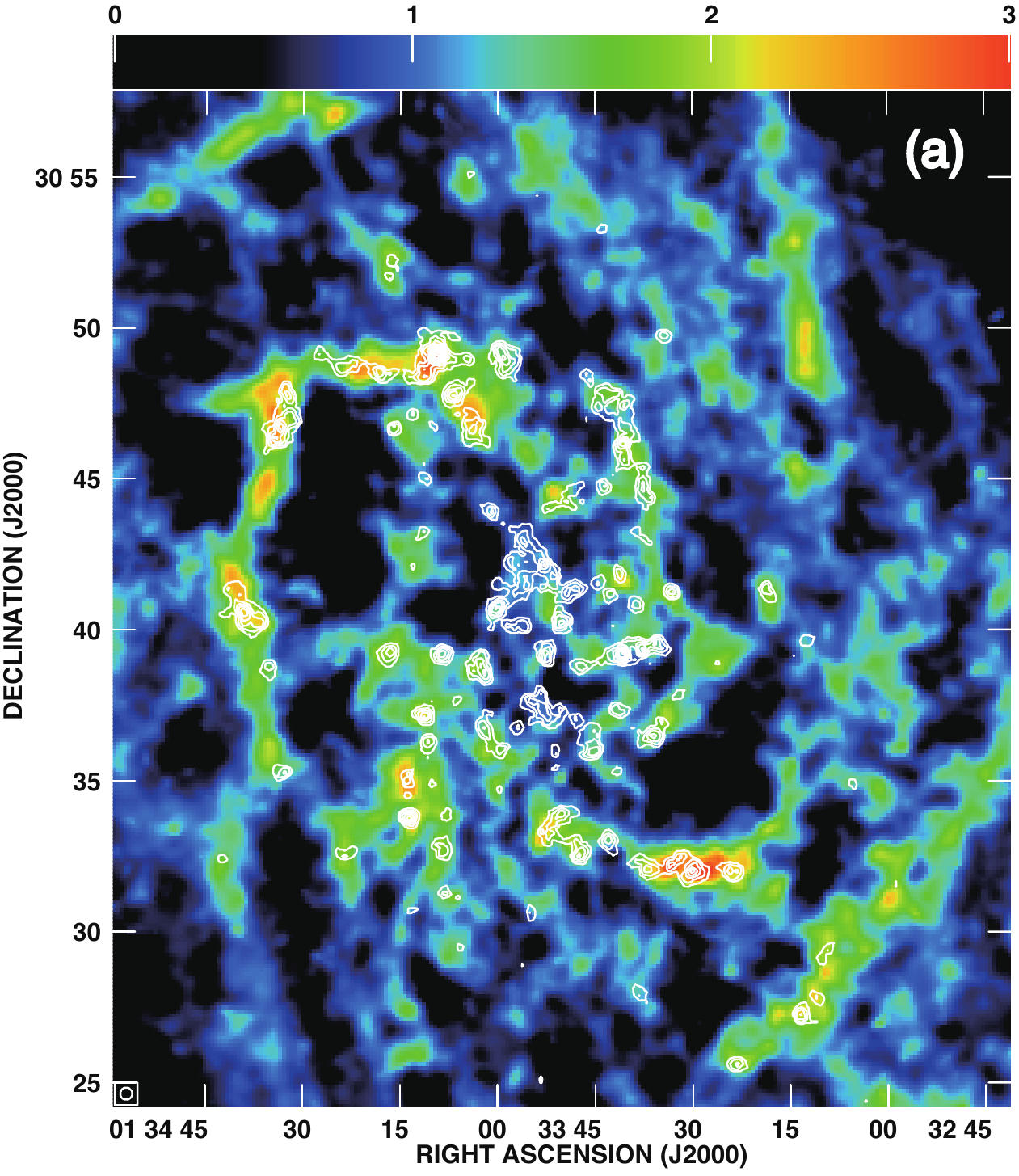}{fig:m33_gmc}{CO(1-0) contours on HI distributions \citep[from ][]{Tosaki:2011fk}.}

\section{GMC Evolution}

Molecular gas dominates even in the inter-arm regions in the MW and M51.
Evidence is being accumulated that the physical conditions of molecular gas
vary systematically as the gas enters spiral arms.

\subsection{M51}

The CO $J=2-1$ and $1-0$ line ratio ($R_{2-1/1-0}$) varies systematically in M51
\citep[][ Figure \ref{fig:co2110} {\it left}]{Koda:2012lr}.
$R_{2-1/1-0}$ rises clearly from a typical low value of $<0.7$ (and often 0.4-0.6) in the
inter-arm regions to a higher value of $>0.7$ (0.8-1.0) in the spiral arms,
particularly at the leading (downstream) edge of the molecular arms.
These high and low $R_{2-1/1-0}$ are similar to those in Galactic GMCs with and without
OB star formation, respectively \citep{Sakamoto:1997ys}.
Thus, the physical conditions of molecular gas evolve during arm passage.

\articlefigure[width=1.0\textwidth]{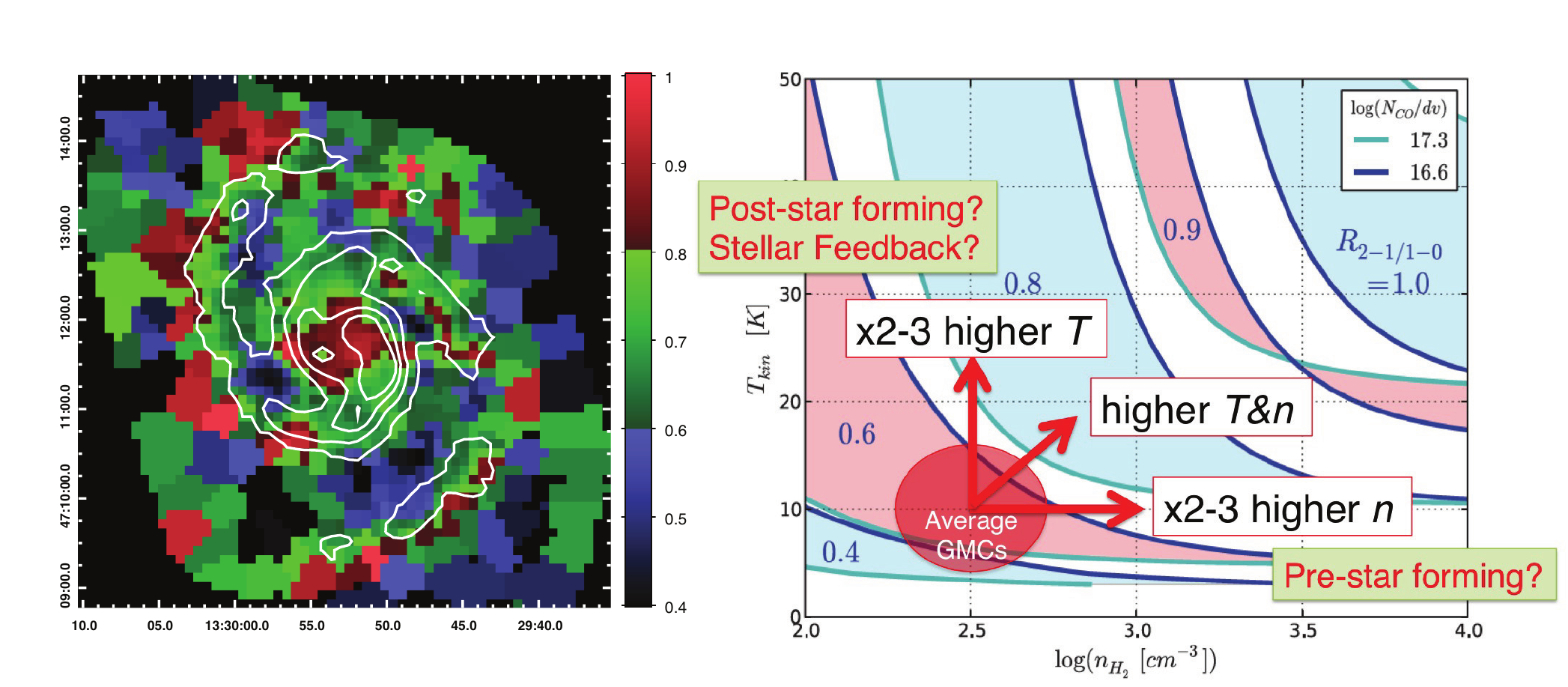}{fig:co2110}{{\it Left:} CO(2-1)/CO(1-0) line ratio map of M51
(applied adaptive smoothing). Contours are CO(1-0). {\it Right:} LVG calculation. From \citet{Koda:2012lr}.}

The current spatial resolution of this image ($\sim1$ kpc) almost certainly blends multiple GMCs.
Assuming all the unresolved GMCs share similar physical conditions,
we can apply a simple one-zone analysis.
Note that virtually all CO emission should come from unresolved GMCs,
since the CO molecule is easily photo-dissociated without self-shielding.
It is also difficult to collisionally excite CO emission outside GMCs, as the critical density
is about the average density within GMCs ($\sim 300\,\rm cm^{-3}$).

A Large Velocity Gradient radiative transfer calculation provides insight into the changes
in the physical conditions. The $R_{2-1/1-0}$  variations indicate that cold and low density
gas ($\lesssim 10$ K, $\lesssim 300$ pc) is required for the interarm GMCs but this gas
must become warmer and/or denser, by a factor of 2-3, in the more active star forming spiral arms
(Figure \ref{fig:co2110} {\it right}).
We cannot separate the increases in temperature and density with the two line analysis.
However, the enhanced
$R_{2-1/1-0}$ at the arm downstream may suggest that the main cause is stellar heating.
Indeed, most star formation is found at the downstream side \citep{Louie:2013lr},
and $R_{2-1/1-0}$ is higher in areas of high $24\mu$m dust surface brightness
(which is an approximate tracer of star formation rate surface density).
Some upstream regions also show high $R_{2-1/1-0}$, possibly indicating density increases
before star formation. The sizes of GMCs need to be shrunk only by 20-30\% as they enter
spiral arms in order to increase GMC densities by a factor of 2-3.

\subsection{The Milky Way}

We can resolve molecular gas structure at small scales in the Milky Way disk and
study its evolution in relation to the spiral structure.

\citet{Sawada:2012pq} demonstrated remarkable variations of
molecular gas structure between inter-arm regions and spiral arms (Figure \ref{fig:maps_armintarm}).
Their line of sight, $l\sim 38\deg$, samples the molecular gas in both the Sagittarius arm
and the inter-arm regions, and hence the regions can be distinguished in velocity channel maps.
The inter-arm emission appears mostly featureless and extended,
while the spiral arm shows a lot of clumps ($\sim 1$ pc in size; either dense or/and warm).
The velocity channel of the highest clump fraction coincides with those of H II regions
and high CO 3-2/1-0 intensity ratio integrated over the field (i.e., warm on average).
It also appears offset from the molecular spiral arm velocity, indicating that the clumps
are at its downstream side.
Therefore, bright and spatially confined structures develop
in a spiral arm, leading to star formation at downstream side, while extended molecular
emission dominates in the inter-arm region.

\articlefigure[width=1.0\textwidth]{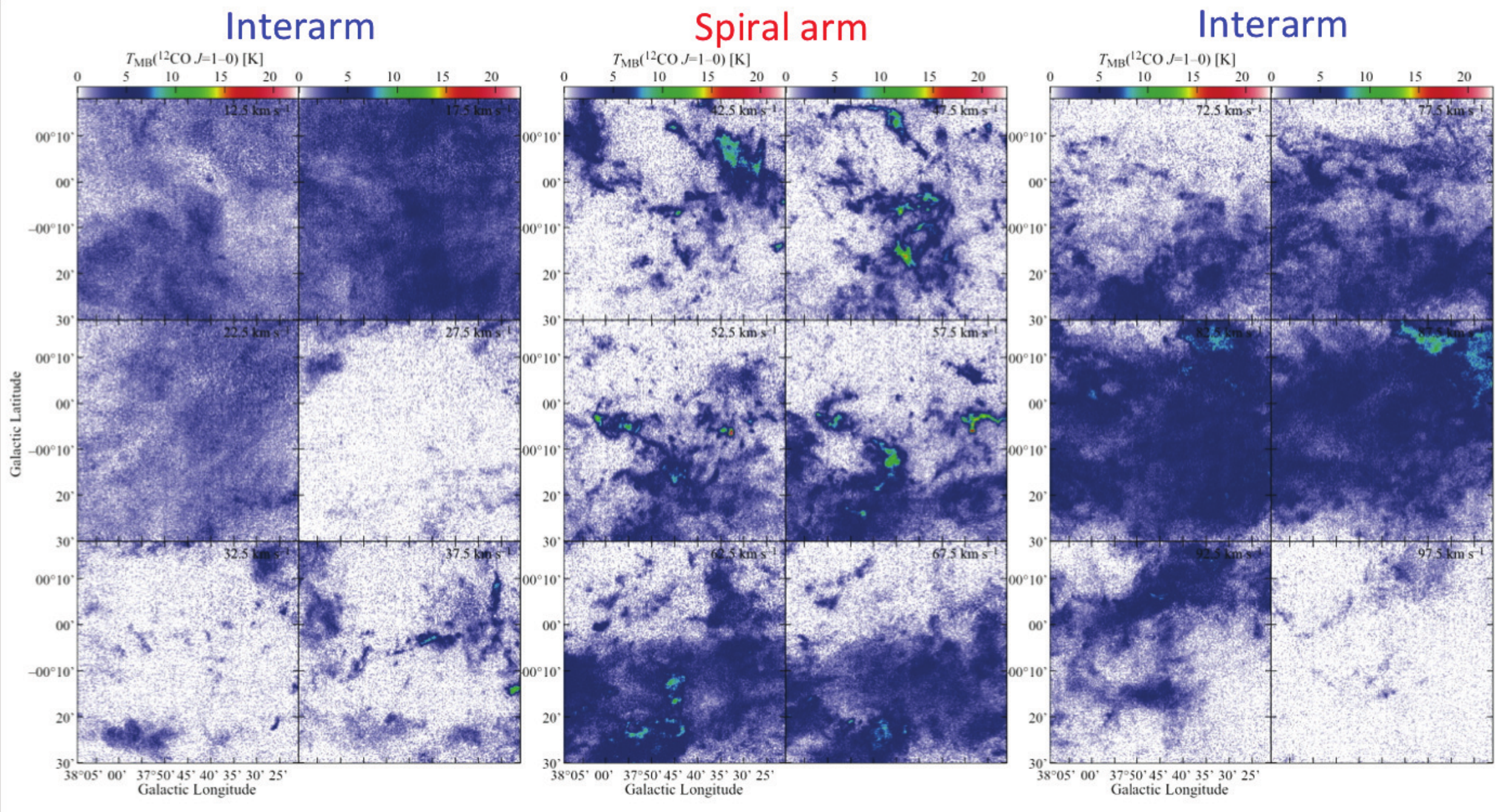}{fig:maps_armintarm}{Spatial distributions of CO(1-0) emission in
inter-arm regions and a spiral arm in the MW \citep{Sawada:2012pq}.}

To quantify the development of the small-scale molecular structure,
\citet{Sawada:2012pq} introduced the brightness distribution index (BDI),
\begin{equation}
BDI = \log_{10} \left( \frac{\int^{T_3}_{T_2} T\cdot B(T)dT}{\int^{T_1}_{T_0} T\cdot B(T)dT} \right),
\end{equation}
with typical boundary brightness temperatures to enclose the extended and lumpy components
($T_0$, $T_1$, $T_2$, $T_3$).
The BDI measures the fractional contribution of spatially confined bright molecular emission over faint emission extended over large areas. This relative quantity is largely independent of the amount of molecular gas and of any conventional, pre-conceived structures, such as cores, clumps, or giant molecular clouds. 

\citet{Sawada:2012lr} confirmed the same evolution over the MW disk.
They applied the BDI to the entire inner MW disk in the northern hemisphere,
using archival data from the Boston University-Five College Radio Astronomy Observatory
$^{13}$CO $J =$ 1-0 Galactic Ring Survey \citep[Figure \ref{fig:bdi}; ][]{Jackson:2006ya}.
The structured molecular gas, traced by higher BDI, appear continuously along the spiral arms
in the $l-v$ diagram.
The high-BDI gas generally coincides with areas with a high population of H II regions,
while there is also some high-BDI gas with no/little signature of ongoing star formation.

These results support the evolutionary sequence
in which unstructured, extended gas transforms itself into a structured state on encountering
the spiral arms, followed by star formation and an eventual return to the unstructured state
after spiral arm passage.

\articlefigure[width=0.9\textwidth]{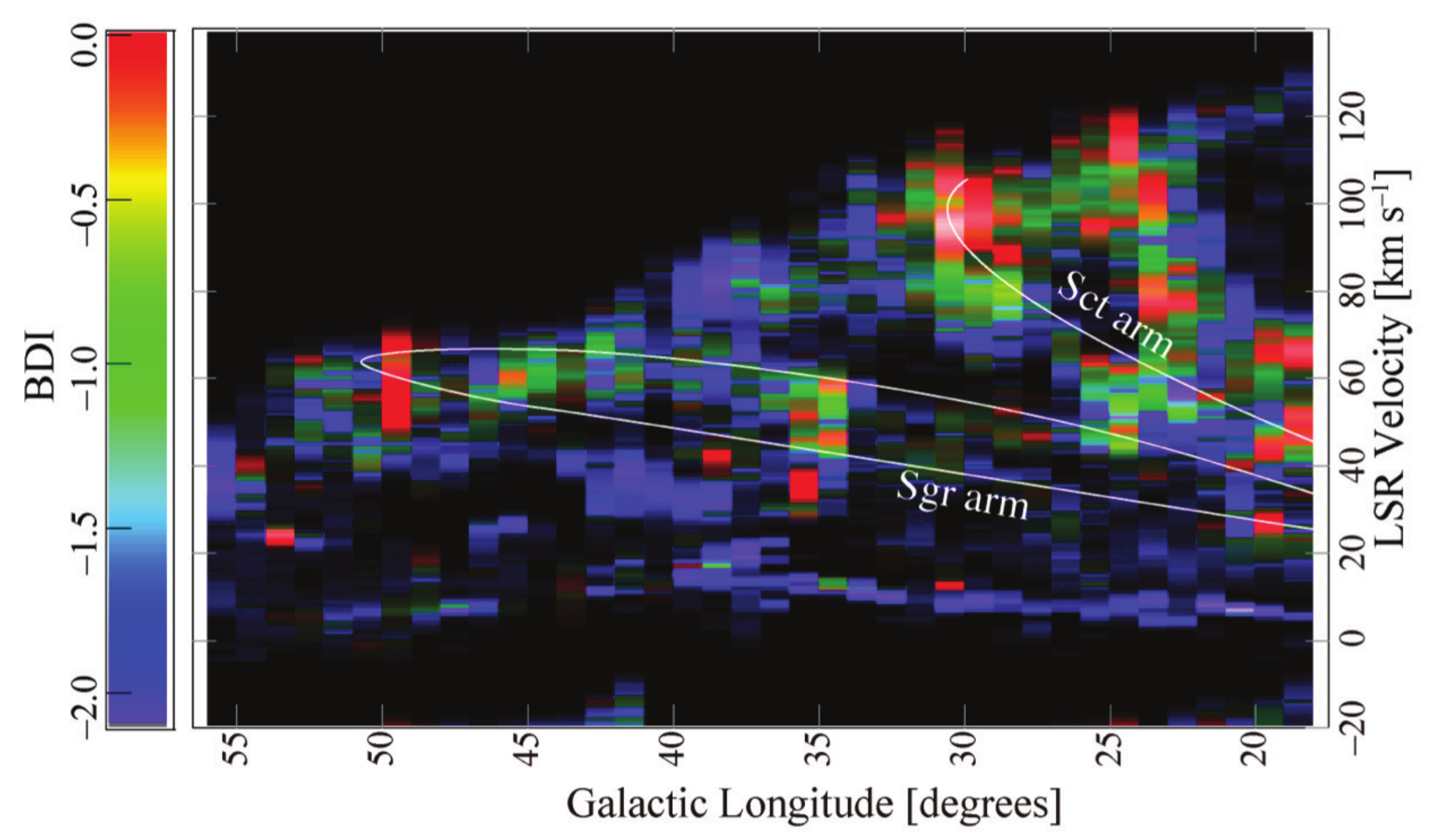}{fig:bdi}{The Brightness Distribution Index (BDI) in the $l-v$ diagram of the MW
\citep{Sawada:2012lr}.}

\section{Summary and Future Prospects}
We saw that the evolution of GMCs depends strongly on the parent galactic environment.
Atom-rich, small (dwarf) galaxies show GMCs almost exclusively on HI spiral arms and filaments,
suggesting their lifetimes as short as an arm-crossing time of 20-30 Myr. On the other hand,
in molecule-rich spiral galaxies GMCs are found almost everywhere, including the inter-arm regions.
The lifetimes there could be as long as a galactic rotation timescale of $\sim 100$ Myr
(they stay molecular throughout galactic rotation, though GMCs can be
coagulated and/or fragmented into next-generation GMCs).
The textbook picture of the gas evolution predicts a rapid atomic/molecular phase change across
spiral arms, but this needs to be revised based on new observations.
We saw clear evidence for the evolution of physical conditions and structure of molecular gas
in galactic disks. The molecular gas and GMCs become denser and/or warmer in spiral arms,
and small pc-scale structures (dense or warm clumps) develop during spiral arm passage.

With ALMA, we are at an exciting moment for furthering understanding of the evolution of the ISM
in galaxies. Extragalactic GMCs will be easily identified and resolved.
Multi-line analyses become possible for individual GMCs with an ALMA sensitivity.
We have not yet seen the full imaging capability of ALMA, as most early observations
are designed solely on sensitivity. Complete synthesis observations with the full ALMA
array, in conjunction with short-spacing data from the total power telescopes and compact 
array, will revolutionize image fidelity, and enable imaging of highly-complex molecular
structures in nearby galaxies.

\acknowledgements JK acknowledges support from the NSF through grant AST-1211680 and NASA through grant NNX09AF40G, a Herschel Space Observatory grant, and an Hubble Space Telescope grant.


\begin{thebibliography}{}
\expandafter\ifx\csname natexlab\endcsname\relax\def\natexlab#1{#1}\fi
\expandafter\ifx\csname url\endcsname\relax
  \def\url#1{\texttt{#1}}\fi
\expandafter\ifx\csname urlprefix\endcsname\relax\def\urlprefix{URL }\fi
\providecommand{\eprint}[2][]{\url{#2}}

\bibitem[{{Cohen} et~al.(1980){Cohen}, {Cong}, {Dame}, \&
  {Thaddeus}}]{Cohen:1980ve}
{Cohen}, R.~S., {Cong}, H., {Dame}, T.~M., \& {Thaddeus}, P. 1980, \apjl, 239,
  L53

\bibitem[{{Dame} et~al.(2001){Dame}, {Hartmann}, \& {Thaddeus}}]{Dame:2001gs}
{Dame}, T.~M., {Hartmann}, D., \& {Thaddeus}, P. 2001, \apj, 547, 792.

\bibitem[{{Engargiola} et~al.(2003){Engargiola}, {Plambeck}, {Rosolowsky}, \&
  {Blitz}}]{Engargiola:2003jo}
{Engargiola}, G., {Plambeck}, R.~L., {Rosolowsky}, E., \& {Blitz}, L. 2003,
  \apjs, 149, 343.

\bibitem[{{Fukui} et~al.(2009){Fukui}, {Kawamura}, {Wong}, {Murai}, {Iritani},
  {Mizuno}, {Mizuno}, {Onishi}, {Hughes}, {Ott}, {Muller}, {Staveley-Smith}, \&
  {Kim}}]{Fukui:2009lr}
{Fukui}, Y., {Kawamura}, A., {Wong}, T., {Murai}, M., {Iritani}, H., {Mizuno},
  N., {Mizuno}, Y., {Onishi}, T., {Hughes}, A., {Ott}, J., {Muller}, E.,
  {Staveley-Smith}, L., \& {Kim}, S. 2009, \apj, 705, 144.

\bibitem[{{Jackson} et~al.(2006){Jackson}, {Rathborne}, {Shah}, {Simon},
  {Bania}, {Clemens}, {Chambers}, {Johnson}, {Dormody}, {Lavoie}, \&
  {Heyer}}]{Jackson:2006ya}
{Jackson}, J.~M., {Rathborne}, J.~M., {Shah}, R.~Y., {Simon}, R., {Bania},
  T.~M., {Clemens}, D.~P., {Chambers}, E.~T., {Johnson}, A.~M., {Dormody}, M.,
  {Lavoie}, R., \& {Heyer}, M.~H. 2006, \apjs, 163, 145.  

\bibitem[{{Kawamura} et~al.(2009){Kawamura}, {Mizuno}, {Minamidani},
  {Filipovi{\'c}}, {Staveley-Smith}, {Kim}, {Mizuno}, {Onishi}, {Mizuno}, \&
  {Fukui}}]{Kawamura:2009lr}
{Kawamura}, A., {Mizuno}, Y., {Minamidani}, T., {Filipovi{\'c}}, M.~D.,
  {Staveley-Smith}, L., {Kim}, S., {Mizuno}, N., {Onishi}, T., {Mizuno}, A., \&
  {Fukui}, Y. 2009, \apjs, 184, 1.

\bibitem[{{Koda} et~al.(2006){Koda}, {Sawada}, {Hasegawa}, \&
  {Scoville}}]{Koda:2006fk}
{Koda}, J., {Sawada}, T., {Hasegawa}, T., \& {Scoville}, N.~Z. 2006, \apj, 638,
  191.

\bibitem[{{Koda} et~al.(2011){Koda}, {Sawada}, {Wright}, {Teuben}, {Corder},
  {Patience}, {Scoville}, {Donovan Meyer}, \& {Egusa}}]{Koda:2011nx}
{Koda}, J., {Sawada}, T., {Wright}, M.~C.~H., {Teuben}, P., {Corder}, S.~A.,
  {Patience}, J., {Scoville}, N., {Donovan Meyer}, J., \& {Egusa}, F. 2011,
  \apjs, 193, 19

\bibitem[{{Koda} et~al.(2012){Koda}, {Scoville}, {Hasegawa}, {Calzetti},
  {Donovan Meyer}, {Egusa}, {Kennicutt}, {Kuno}, {Louie}, {Momose}, {Sawada},
  {Sorai}, \& {Umei}}]{Koda:2012lr}
{Koda}, J., {Scoville}, N., {Hasegawa}, T., {Calzetti}, D., {Donovan Meyer},
  J., {Egusa}, F., {Kennicutt}, R., {Kuno}, N., {Louie}, M., {Momose}, R.,
  {Sawada}, T., {Sorai}, K., \& {Umei}, M. 2012, \apj, 761, 41.  

\bibitem[{{Koda} et~al.(2009){Koda}, {Scoville}, {Sawada}, {La Vigne}, {Vogel},
  {Potts}, {Carpenter}, {Corder}, {Wright}, {White}, {Zauderer}, {Patience},
  {Sargent}, {Bock}, {Hawkins}, {Hodges}, {Kemball}, {Lamb}, {Plambeck},
  {Pound}, {Scott}, {Teuben}, \& {Woody}}]{Koda:2009wd}
{Koda}, J., {Scoville}, N., {Sawada}, T., {La Vigne}, M.~A., {Vogel}, S.~N.,
  {Potts}, A.~E., {Carpenter}, J.~M., {Corder}, S.~A., {Wright}, M.~C.~H.,
  {White}, S.~M., {Zauderer}, B.~A., {Patience}, J., {Sargent}, A.~I., {Bock},
  D.~C.~J., {Hawkins}, D., {Hodges}, M., {Kemball}, A., {Lamb}, J.~W.,
  {Plambeck}, R.~L., {Pound}, M.~W., {Scott}, S.~L., {Teuben}, P., \& {Woody},
  D.~P. 2009, \apjl, 700, L132.

\bibitem[{{Louie} et~al.(2013){Louie}, {Koda}, \& {Egusa}}]{Louie:2013lr}
{Louie}, M., {Koda}, J., \& {Egusa}, F. 2013, \apj, 763, 94.

\bibitem[{{Mac Low}(1999)}]{Mac-Low:1999rw}
{Mac Low}, M.-M. 1999, \apj, 524, 169.

\bibitem[{{Miura} et~al.(2012){Miura}, {Kohno}, {Tosaki}, {Espada}, {Hwang},
  {Kuno}, {Okumura}, {Hirota}, {Muraoka}, {Onodera}, {Minamidani}, {Komugi},
  {Nakanishi}, {Sawada}, {Kaneko}, \& {Kawabe}}]{Miura:2012yq}
{Miura}, R.~E., {Kohno}, K., {Tosaki}, T., {Espada}, D., {Hwang}, N., {Kuno},
  N., {Okumura}, S.~K., {Hirota}, A., {Muraoka}, K., {Onodera}, S.,
  {Minamidani}, T., {Komugi}, S., {Nakanishi}, K., {Sawada}, T., {Kaneko}, H.,
  \& {Kawabe}, R. 2012, \apj, 761, 37.

\bibitem[{{Rand} \& {Kulkarni}(1990)}]{Rand:1990fb}
{Rand}, R.~J., \& {Kulkarni}, S.~R. 1990, \apjl, 349, L43

\bibitem[{{Sakamoto} et~al.(1997){Sakamoto}, {Hasegawa}, {Handa}, {Hayashi}, \&
  {Oka}}]{Sakamoto:1997ys}
{Sakamoto}, S., {Hasegawa}, T., {Handa}, T., {Hayashi}, M., \& {Oka}, T. 1997,
  \apj, 486, 276

\bibitem[{{Sanders} et~al.(1985){Sanders}, {Scoville}, \&
  {Solomon}}]{Sanders:1985ud}
{Sanders}, D.~B., {Scoville}, N.~Z., \& {Solomon}, P.~M. 1985, \apj, 289, 373

\bibitem[{{Sawada} et~al.(2012{\natexlab{a}}){Sawada}, {Hasegawa}, \&
  {Koda}}]{Sawada:2012lr}
{Sawada}, T., {Hasegawa}, T., \& {Koda}, J. 2012{\natexlab{a}}, \apjl, 759,
  L26.

\bibitem[{{Sawada} et~al.(2012{\natexlab{b}}){Sawada}, {Hasegawa}, {Sugimoto},
  {Koda}, \& {Handa}}]{Sawada:2012pq}
{Sawada}, T., {Hasegawa}, T., {Sugimoto}, M., {Koda}, J., \& {Handa}, T.
  2012{\natexlab{b}}, \apj, 752, 118.

\bibitem[{{Scoville} \& {Wilson}(2004)}]{Scoville:2004lr}
{Scoville}, N.~Z., \& {Wilson}, C.~D. 2004, in The Formation and Evolution of
  Massive Young Star Clusters, edited by H.~J.~G.~L.~M. {Lamers}, L.~J.
  {Smith}, \& A.~{Nota}, vol. 322 of Astronomical Society of the Pacific
  Conference Series, 245

\bibitem[{{Tosaki} et~al.(2011){Tosaki}, {Kuno}, {Onodera}, {Sawada},
  {Muraoka}, {Nakanishi}, {Komugi}, {Nakanishi}, {Kaneko}, {Hirota}, {Kohno},
  \& {Kawabe}}]{Tosaki:2011fk}
{Tosaki}, T., {Kuno}, N., {Onodera}, S.~M., Rie, {Sawada}, T., {Muraoka}, K.,
  {Nakanishi}, K., {Komugi}, S., {Nakanishi}, H., {Kaneko}, H., {Hirota}, A.,
  {Kohno}, K., \& {Kawabe}, R. 2011, \pasj, 63, 1171.

\end{thebibliography}
\end{document}